\documentclass{article}

\usepackage[dvips]{graphicx}
\usepackage{psfrag}
\usepackage{natbib}

\usepackage{amsmath}

\newcommand{\degree}{\ensuremath{^\circ}}
\newcommand{\apj}{Astroph. J.}
\newcommand{\solphys}{Sol. Phys.}
\newcommand{\apjs}{Astroph. J. Suppl.}

\usepackage{epsfig}

\author{T.  Hartlep$^a$, A. G. Kosovichev$^a$, J. Zhao$^a$ \& N. N. Mansour$^b$}

\date{{\sl $^a$W. W. Hansen Experimental Physics Laboratory, \\ Stanford University, Stanford, CA \\
$^b$NASA Ames Research Center, Moffett Field, CA}}

\title{Signatures of emerging \\ subsurface structures in the sun}  

\begin{document}

\maketitle

\section{Motivation and objectives}

The complex dynamics that lead to the emergence of active regions on the sun are poorly understood. 
One possibility is that magnetic structures (flux tubes, etc.) rise from below the surface by self induction and convection that lead to the formation of active regions and sunspots on the solar surface. For space weather forecasting, one would like to detect the subsurface structures before they reach the surface. The goal of this study is to investigate whether sound speed perturbations associated with subsurface structures could affect the acoustic power observed at the solar surface above them. Possible mechanisms for this effect are wave reflection, scattering or diffraction. 

By using numerical simulations of wave propagation in the solar interior, we investigate whether observations of the acoustic power can be used to detect emerging active regions before they appear on the surface.
In the simulations, subsurface structures are modeled as regions with enhanced or reduced acoustic wavespeed.
We show how the acoustic power above a subsurface region depends on the sign, depth and strength of the wavespeed perturbation.
For comparison, we analyze observations from {\sl SOHO}/MDI of the emergence of solar active region NOAA~10488.

\section{Numerical method}

\subsection{Simulation code}

In the following, we give a brief overview of the numerical simulation code used in this study.
For more details, see \citet{har08} and \cite{har05}.

We model solar acoustic oscillations in a spherical domain by using linearized Euler equations and consider a static background in which localized variations of the sound speed are taken into account.
Simulating the 3-D wavefield in the full solar interior is challenging, and many simplifications are needed to make such simulations feasible.
The oscillations are assumed to be adiabatic, and are driven by randomly forcing density perturbations near the surface.
For the unperturbed background model of the sun, we use the standard solar model S of~\citet{chr96} matched to a model for the chromosphere~\citep{ver81}.
Localized sound speed perturbations of various sizes are added as simple subsurface structures.
Non-reflecting boundary conditions are applied at the upper boundary by means of an absorbing buffer layer with a damping coefficient that is set to zero in the interior and increases smoothly into the buffer layer.

The linearized Euler equations describing wave propagation in the sun are written in the form:
\begin{eqnarray}
   \label{Eq:E1}          
   \partial_t \rho^\prime & = & - \Phi^\prime + S - \chi \rho^\prime,  \\
   \label{Eq:E2 }
   \partial_t \Phi^\prime & = & - \Delta c^2 \rho^\prime + \nabla \cdot \rho^\prime \mathbf{g_0} - \chi \Phi^\prime,
 \end{eqnarray}
where $\rho^\prime$ and $\Phi^\prime$ are the density perturbations and the divergence of the momentum perturbations associated with the waves, respectively. Quantity $S$ is a random function mimicking acoustic sources, $c$ is the background sound speed, $\mathbf{g_0}$ is the acceleration due to gravity, and $\chi$ is the damping coefficient used in the absorbing buffer layer.
Perturbations of the gravitational potential are neglected, and the adiabatic approximation is used.
In order to make the linearized equations convectively stable, we also neglect the entropy gradient of the background model.
The calculations show that this assumption does not significantly change the propagation properties of acoustic waves including their frequencies, except for the acoustic cut-off frequency, which is slightly reduced.
For comparison, other authors have modified the solar model including its sound speed profile~\citep[e.g.,][]{hanasoge2006,par07} in order to stabilized their simulations.
In those cases, the oscillation mode frequencies may differ significantly from the frequencies observed for the real sun.

For numerical discretization, a Galerkin scheme is applied where spherical harmonic functions are used for the angular dependencies, and fourth order B-splines~\citep{lou97,kra99} are used for the radial direction;
Two-thirds dealiasing is used when computing the $c_0^2\rho^\prime$-term in angular space. All other operations are performed directly in spherical harmonic coefficient space.
The radial resolution of the B-spline method is varied proportionally to the speed of sound, i.e., the generating knot points are closely spaced near the surface (where the sound speed is small), and are coarsely spaced in the deep interior (where the sound speed is large). The simulations presented in this paper employ the spherical harmonics of angular degree $l$ from 0 to 170, and a total of 300 B-splines in the radial direction.
A staggered Yee scheme is used for time integration, with a time step of 2~seconds.

\begin{figure}
  \centering
  \vspace{5mm}
  \includegraphics[width=5.5cm]{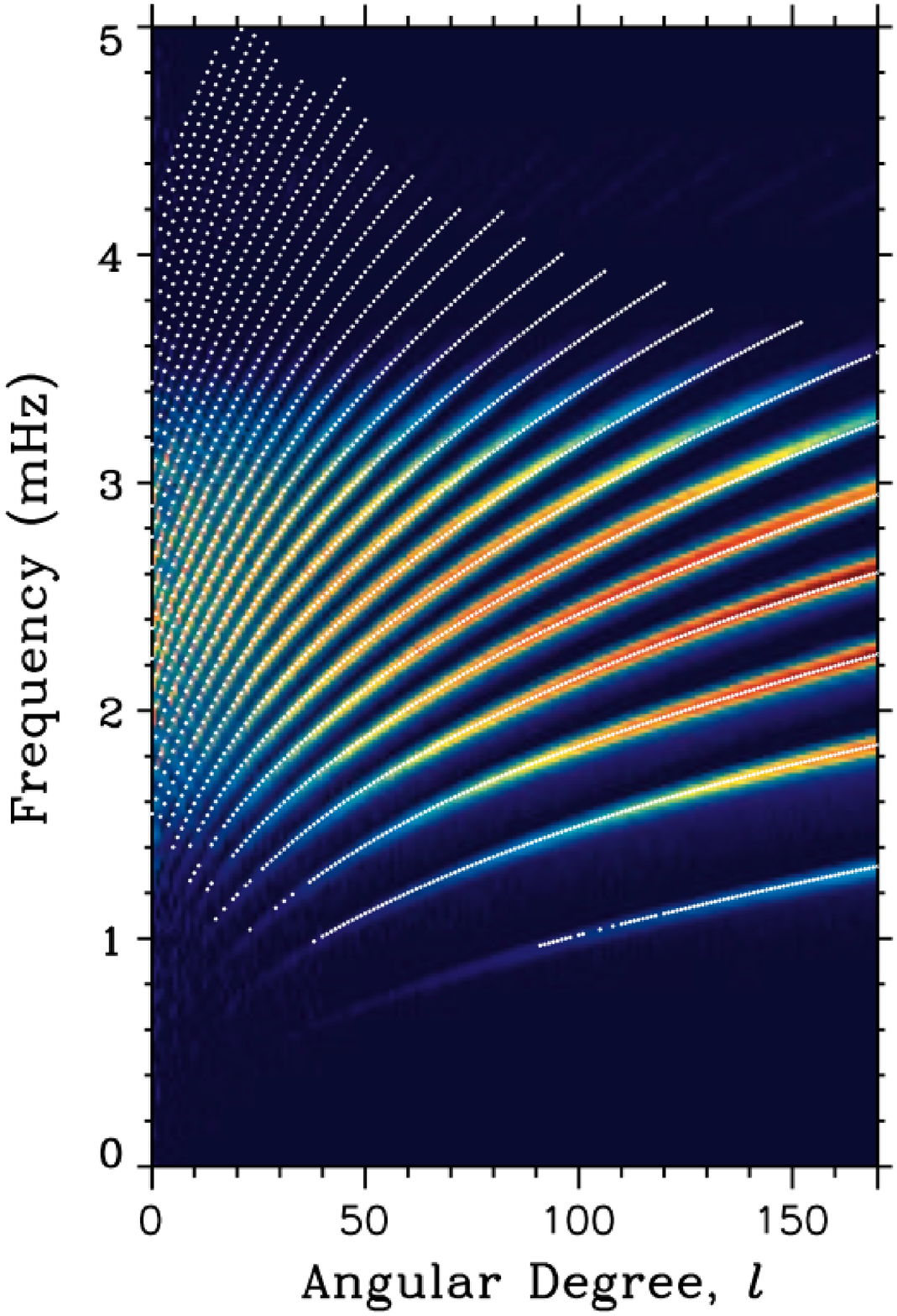}
  \vspace{5mm}
  \caption{Power spectrum of the radial velocity at 300~km above the photosphere computed from a simulated dataset. For comparison, white dots indicate the observed frequencies obtained from 144 days of MDI medium-$l$ data using the averaged-spectrum method~\citep{rho97}.}
  \label{Fig:Power}
\end{figure}

The simulation code has been successfully used for testing helioseismic far-side imaging by simulating the effects of model sunspots on the acoustic field~\citep{har08}.
A power spectrum diagram demonstrating good agreement between the simulation and the oscillation frequencies derived from solar observations
was presented in~\citet{har08} and is reproduced here in Fig.~\ref{Fig:Power}.
As previously noted, the model has a slightly lower cut-off frequency than the real sun.
Because of this, high mode frequencies above approximately 4~mHz seen in observations are not present in the simulation.

\subsection{Subsurface structures}

For this study, we consider simple subsurface structures in which the sound speed $c$ differs from the sound speed of the standard solar model, $c_o$, in the following fashion:
\begin{equation}
  \left( \frac{c}{c_o} \right)^2 = 1 + A
    \begin{cases}
      \frac{1}{4} \left( 1+\cos \pi \frac{z-z_s}{d_s}  \right)  \left( 1 + \cos \pi \frac{\alpha}{\alpha_d} \right) & \mbox{for~} |\frac{\alpha}{\alpha_d}| \le 1, |\frac{z-z_s}{d_s}| \le 1; \\
      0 & \mbox{otherwise},
    \end{cases}
\end{equation}  
where $\alpha$ is the angular distance from the center of a subsurface structure, and $z$ is the depth from the photosphere.
The quantities $z_s$, $d_s$, $\alpha_d$ and $A$ denote the depth at which the sound speed perturbation is largest, the radial and angular sizes, and the amplitude of the sound speed variation, respectively.

\section{Results}

\subsection{Simulations}

\begin{figure}
  \centering
  \vspace*{5mm}
  \vspace*{7mm}\includegraphics[width=11.0cm]{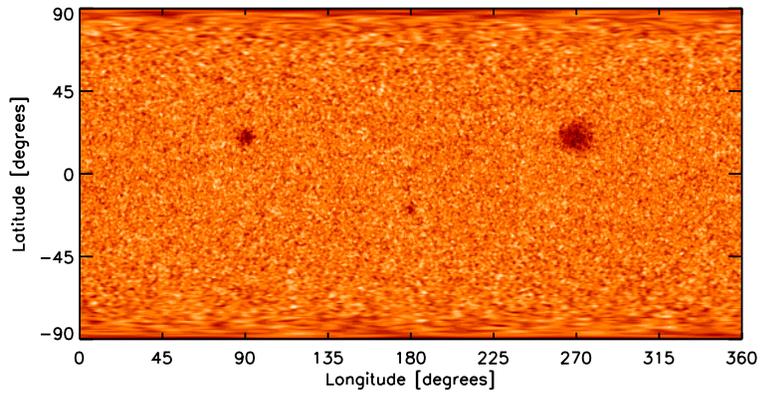}
  \vspace*{8mm}
  \caption{Acoustic power map at 300~km above the photosphere from a simulation with three different subsurface regions each with a maximum of 10\% reduction in sound speed at a depth of 20~Mm. The quantity shown is the square of the radial velocity, averaged over 824~minutes, with low and high values indicated in dark and bright, respectively. A model subsurface region of 45~Mm radius is located at a latitude of $-20\degree$ and longitude of $180\degree$. Two larger regions of 90 and 180~Mm radius are located at a latitude of $+20\degree$, and longitudes of $90\degree$ and $270\degree$, respectively.}
  \vspace{4mm}
  \label{Fig:Powermap}
\end{figure}

Simulations have been performed for subsurface regions of three angular sizes $\alpha_d=3.7\degree, 7.4\degree$ and $14.8\degree$ corresponding to radii at the solar surface of 45, 90 and 180~Mm, respectively.
Two depths, 20 and 30~Mm, and three amplitudes were considered.
The radial size $d_s$ was 20~Mm for all cases.
To reduce the number of separate calculations, each simulation contained three subsurface regions, which were placed far enough apart to avoid interferences.
Figure~\ref{Fig:Powermap} shows an example of an acoustic power map obtained from one of the simulations. In this case, the three subsurface regions have reduced sound speed. A significant reduction in the acoustic power above all three regions can be seen. 

\begin{figure}
	\centering
         \vspace*{3mm}
	\hspace*{-5mm}\includegraphics[width=8.5cm]{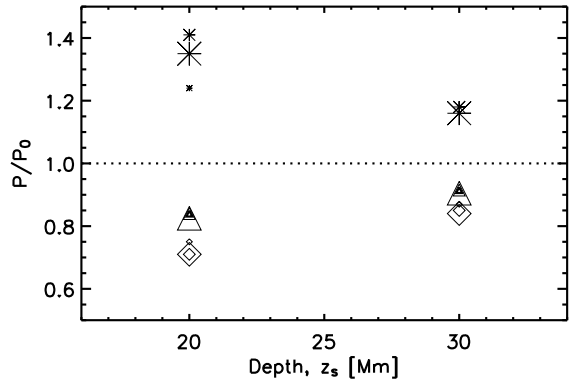}
         \vspace*{2mm}
	\caption{Ratio between the average acoustic power in the center above a subsurface region and away from the region as a function of the depth location of the perturbation. The symbols denote subsurface regions of various strengths: (stars) 10\% sound speed increase,  (triangles) 5\% decrease and (diamonds) 10\% decrease. The region sizes, 45, 90 and 180~Mm radius, are indicated by differently sized symbols from small to large, respectively.}
         \vspace{3mm}
	\label{Fig:Simulation:PowerVsDepth}
\end{figure}

A compilation of the main simulation results is given in Fig.~\ref{Fig:Simulation:PowerVsDepth}.
The diagram shows the ratio between the acoustic power in the center above a subsurface region and far away from such regions (in a ``quiet sun'' zone).
The acoustic power is measured by averaging the square of the radial velocity.
The horizontal velocity components of the waves considered here (low to medium spherical harmonic degrees) can be neglected at the solar surface.
It is evident that regions with positive sound speed perturbation cause an increase in the acoustic power, and that negative perturbations cause a decrease. 
Also, the stronger sound speed perturbations produce stronger power changes.
Except for one case, the maximum power change depends only weakly on the size of the subsurface region and is generally larger for the larger regions.
The power change is weaker for deeper perturbations.
However, in all our cases the power change is close to or greater than 10\%.
Thus, if such structures exist, they should be measurable from relatively short time series of observations of solar oscillations.

\subsection{Observations}

\begin{figure}
	\centering
         \vspace*{6mm}
	\hspace*{-3mm}\includegraphics[width=12.0cm]{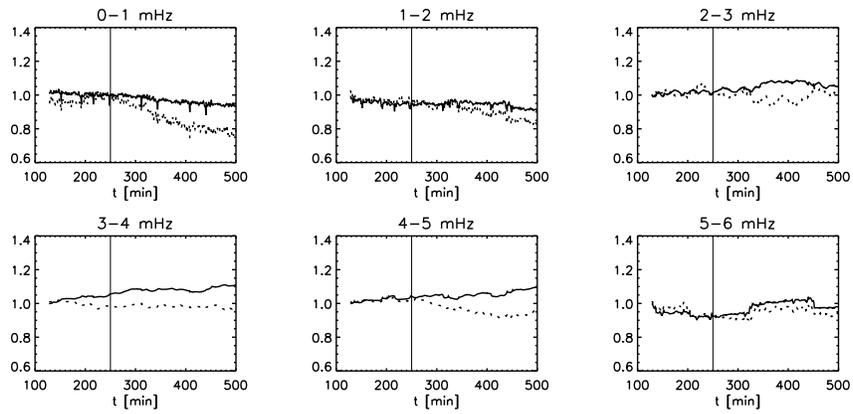}
         \vspace{0.5mm}
	\caption{Time series of the observed acoustic power during the emergence of active region NOAA~10488 on October 26, 2003. Time $t$ starts from 06$^h$00$^m$00$^s$ UT. Shown is the power density of Doppler velocity increments (1-minute cadence) inside a patch where the active region is emerging (dotted lines) and outside the patch (solid lines) averaged over the preceding 128~minutes, each divided by the initial power density outside. Except for the splitting into frequency intervals, no filtering was done. Vertical lines indicate the time when significant magnetic flux has emerged.}
	\label{Fig:Observation:TimeSeries}
\end{figure}

Encouraged by the simulation results, we have tried to find evidence for this effect in observations.
Figure~\ref{Fig:Observation:TimeSeries} shows the result from our analysis of the emergence of active region NOAA~10488.
The observational data is split into six 1~mHz wide frequency intervals, since the different frequencies show different behaviors. 
In most of the frequency intervals, the power inside the patch where the region emerges and outside was very much the same until about the 250th~minute of the time series starting at 06$^h$00$^m$00$^s$ UT, October 26, 2003.
This is the time when significant magnetic flux was first seen on the surface, although some magnetic field was visible already at about the 220th~minute. 
The only frequency interval that shows a significant difference in acoustic power before the emergence is from 3 to 4~mHz.
Here, the two curves for the area of the emerging active region and the quiet sun diverge about 100~minutes before significant magnetic flux appears on the surface.
The power inside the patch was lower than outside, possibly indicating a subsurface region with reduced wavespeed (see Fig.~\ref{Fig:Simulation:PowerVsDepth}).
The reason why only a signal is seen in that frequency interval is currently unknown.

\section{Conclusions and future work}
By using numerical simulations of acoustic wave propagation, we have shown that under suitable conditions, subsurface structures in the sun can alter the average acoustic power observed at the solar photosphere.
This effect could potentially be used for detecting emerging subsurface structures before they reach the surface.
We have also analyzed observational data of the emergence of solar active region NOAA~10488, and have demonstrated a reduction in the observed acoustic power before the emergence.
In the future, we plan to analyze data from other emergence events for comparison.




\clearpage

\clearpage

\clearpage

\clearpage

\end{document}